# Radio-Transparent Dipole Antenna Based on a Metasurface Cloak

Jason Soric[1,*], Younes Ra'di[2,*], Diego Farfan[1,2], and Andrea Alù[1,2,3,4]

[1]Department of Electrical and Computer Engineering, The University of Texas at Austin, Austin, TX 78712, USA
[2]Photonics Initiative, Advanced Science Research Center, City University of New York, NY 10031, USA
[3]Physics Program, Graduate Center, City University of New York, NY 10016, USA
[4]Department of Electrical Engineering, City College of The City University of New York, NY 10031, USA

*These authors contributed equally to this work.

To whom correspondence should be addressed: aalu@gc.cuny.edu

**Antenna technology is at the basis of ubiquitous wireless communication systems and sensors. Radiation is typically sustained by conduction currents flowing around resonant metallic objects that are optimized to enhance efficiency and bandwidth. However, resonant conductors are prone to large scattering of impinging waves, leading to challenges in crowded antenna environments due to blockage and distortion. Metasurface cloaks have been explored in the quest of addressing this challenge by reducing antenna scattering, but with limited performance in terms of bandwidth, footprint and overall scattering reduction. Here we introduce a different route towards radio-transparent antennas, in which the cloak itself acts as the radiating element, drastically reducing the overall footprint while enhancing scattering suppression and bandwidth, without sacrificing other relevant radiation metrics compared to conventional antennas. This technique offers a new application of cloaking technology, with promising features for crowded wireless communication platforms and noninvasive sensing.**

High-density integrated wireless systems continue to grow at a very fast pace in our ever-connected world. Driven by increasingly overcrowded wireless channels, there is a burgeoning demand for co-located antennas and sensors operating in overlapping or disjoint frequency channels on the same physical platform. However, nearby antennas act as parasitic elements for each other, with negative impact on each other's radiation performance, including beam-squinting, blockage (shadowing), and gain distortion of neighboring radiation patterns. The parasitic effects concurrently act as detuning elements on the mutual input impedance [1]. A technique based on choked antenna arms has been explored to address these challenges in interleaved dual-band base station antenna arrays [2], but with inherent limitations on the antenna shape and performance.

In an effort to address these challenges, metasurfaces and metamaterials [3]-[14] have been explored as a way to eliminate antenna blockage and co-site interference reduction in a compact and scalable platform, exploiting the concept of cloaking and scattering cancellation [15]-[22]. In particular, metasurface cloaks based on hard surfaces [23]-[24], transmission lines [25]-[27], and scattering cancellation techniques [28]-[41] have been applied over the years to reduce antenna blockage in various scenarios. These approaches leverage the effectiveness of metasurface cloaks to suppress the scattering and shadows of conventional conductive targets, restoring the impinging wavefronts independent of the source location. However, resonant antennas have very strong scattering to start with over the entire frequency band of interest, making the goal of large scattering suppression, at the same time avoiding detuning, a challenging task. An additional significant disadvantage of these methods is the increased overall size of the cloaked antenna and its fabrication complexity. In particular, given that conventional antennas are conductive, a conventional metasurface cloak cannot be simply wrapped around the metal, but it needs to include a spacer with sufficient thickness dictated by the required bandwidth [32]. This feature leads to awkward designs that may not be easily scaled for mass production. The large footprint of metasurface cloaks over conventional antenna systems also implies polarization selectivity, effective operation over a limited range of incidence angles, and narrow operational bandwidth. These restrictions have so far hindered the broad practical impact of cloaking technology in wireless antenna and sensor systems.

Here, we explore an alternative route to enable radio-transparent antenna technology, which addresses all these challenges. We start from a metasurface cloak designed to suppress the scattering from a dielectric core. Given the limited scattering of nonresonant dielectric objects of subwavelength size, the cloak can make the overall scattering very low over a broad bandwidth. At the same time, given the non-conductive nature of the object, the cloak can be wrapped conformally to the object, or even embedded inside it, providing broadband scattering suppression and angular stability in a very compact design. Quite interestingly, the cloaking surface aimed at canceling the scattering of the dielectric core can actually be designed to also operate as an antenna over the desired frequency band. We demonstrate its operation in a commercial base station antenna panel, showing that it can meet all relevant metrics of performance in terms of radiation features, enabling an ideally suited broadband radio-transparent antenna.

In order to showcase its operation, we consider a typical wireless communication scenario, in which antennas operating in different frequency bands are forced to operate in close proximity. A simplified illustrative example, consistent with the typical geometry of a multi-band radio-base station antenna array, is shown in Fig. 1, where two dipole antennas of different size operating in different frequency bands are located close to each other. The smaller antenna operates in the higher frequency band ($f_{HB}$) and it does not significantly affect the radiation features of the larger antenna operating in the lower band ($f_{LB}$) given the small size. However, the larger antenna drastically deteriorates the radiation at $f_{HB}$, affecting its polarization, beam-squinting, and inducing undesired shadows in the radiation pattern [see Fig. 1(a)]. Our goal is to address this challenge by realizing a radio-transparent low-frequency antenna that offers minimal interference [see Fig. 1(b)]. Compared to other approaches, we demonstrate a broad bandwidth over which the antenna has very limited scattering signatures, with a design that can be straightforwardly extended to different antenna lengths and complex geometries within a simple and scalable fabrication process.

Indeed, conventional cloaking approaches have been explored to suppress the scattering from low-frequency antennas in a similar configuration over the entire high-frequency band. However, several issues limit the applicability of conventional cloaking technology to resonant conducting antennas. The most important challenge is the tradeoff between operational bandwidth and cloak thickness: because of the conductive nature of the antenna to be cloaked, a too narrow gap between metasurface cloak and antenna implies a large sensitivity to the operation frequency, making the cloak ineffective. At the same time, thicker cloaks are not only undesirable due to footprint considerations, but their response also becomes angular dependent, making the scattering reduction very sensitive to the location of the neighboring antennas. As an illustrative example, Figs. 2(a) and 2(b) show the normalized scattering width as a function of frequency and angle for a metallic cylinder cloaked with an optimal metasurface varying the distance from the cylinder surface [designs shown in the insets]. Since we aim at suppressing the influence on high-frequency antennas placed in the near-field, the scattering suppression needs to be effective over a wide range of incidence angles. As expected, a wider frequency bandwidth is possible only at the cost of thicker designs, producing increased sensitivity to the illumination angle [see Fig. 2(c)]. The reason for this trade-off lies in the fact that the object we are trying to cloak is a strong scatterer to start with, typically a conducting dipole designed to efficiently radiate at a nearby frequency from the one in which we aim at suppressing the scattering.

To address this issue, we exploit a conceptually different approach, designing our metasurface for a subwavelength dielectric scatterer. The inset in Fig. 2(d) shows the schematic of our cloaked object, consisting of a dielectric hollow core replacing the high-scattering metallic cylinder. A suitably tailored cloaking metasurface, inductive in nature, is inserted within the dielectric object to suppress its (capacitive) scattering, offering a dramatic improvement in terms of frequency and angular bandwidth, as seen in Fig. 2(d). Particularly for larger angles, the dielectric object has very low scattering to start with, so the metasurface only needs to suppress the residual scattering for close-to-normal incidence, a much more relaxed goal than in the previous design. The angular sensitivity is hence drastically reduced [42]. At the same time, the implemented metasurface cloak, due to its conducting nature, can be optimized to be impedance matching to a regular feeding line and used itself as a radiating element. In this configuration, the dielectric core thus serves three purposes: it acts as a contrast element to eliminate the scattering of the metasurface, it offers mechanical support, and it capacitively enhances the coupling of the metasurface with the feeding source.

In Fig. 3 we study the cloaking and radiation features of our radio-transparent antenna based on this concept. Shown in the left inset of Fig. 3 is the design concept. The structure was optimized, as detailed in the Supplementary Material, to operate in the 698-968 MHz (LB) and suppress the scattering in the 1.71-2.710 GHz (HB) range, since these bands are used for 3G and 4G LTE services. The frequency range around 3.5 GHz (UB) is also highlighted, as it is relevant in small cell or shared spectrum communications to alleviate the ever-growing wireless demands. The scattering features of our antenna are dominated by the electric dipole contribution, maximally excited by transverse magnetic (TM)-polarized waves at normal incidence, hence in this figure we analyze the scattering cross section for this excitation. The response for different incidence angles and different polarizations is shown in the Supplementary Material, featuring negligible scattering compared to Fig. 3. The figure compares the total scattering cross section (SCS) of a conventional conducting dipole, of a dielectric cylinder, and of our radio-transparent antenna, all with same radii and length. The conventional dipole antenna, whose geometry is optimized to efficiently radiate in the LB, shows a significant SCS across the HB and UB, with a self-resonance around 1.5 GHz. The bare low-loss host dielectric (without embedded inductive cloak) with $\varepsilon = \varepsilon_r \left(1 - j \tan \delta \right)$, where $\varepsilon_r = 4.4$ and $\tan \delta = 0.0005$, has a much lower scattering, which grows in the upper frequency range as expected. The cloaked cylinder [28]-[31] is very effective at suppressing the scattering in the HB and UB bands, and it also offers a pronounced scattering dip centered in the LB band, of great interest to eliminate scattering interference with surrounding LB antennas in the array. The metasurface not only suppresses the scattering across the frequency range of interest compared to the dielectric rod (and much more if we compare it with the original conducting dipole), but it also offers a means to guide conduction currents that can radiate in the LB frequency range. It is also important to stress that the very large bandwidth over which the scattering is suppressed in this design is enabled over an ultracompact footprint, with the cloak actually embedded within the dielectric object, therefore with zero contributions to the overall footprint of the antenna. This is very different than any other approach to scattering suppression, which are typically unfeasible in terms of the resulting footprint, and lead to increased scattering for TE-polarized wavefronts and other incidence angles due to the increased geometrical cross-section [25],[33].

We now test the performance of our optimized cloak in terms of radiation features, exploring its LB performance in a conventional 3G and 4G LTE radio-base station, as shown in Fig. 4(a). The figure compares three scenarios of interest: an isolated arrays of HB crossed dipoles, placed at a quarter wavelength from the ground plane; the dual-band array in which LB crossed dipoles are placed on top of the HB array, again at a quarter wavelength from the ground plane for their longer radiation wavelengths, and finally our radio-transparent antenna in a cross-dipole configuration replacing the standard LB dipole. The LB elements are placed directly in front of the HB array because of the required distance from the ground, typically introducing a large disturbance in terms of beam squint, which re-directs the HB radiation pattern away from its intended boresight direction $\left(\theta_i = 90°\right)$. The beamwidth and gain of the HB elements can also be significantly altered by the presence of the LB elements. The total SCS of the LB element (conductive dipole antenna), as plotted in Fig. 3, is a good measure of its electromagnetic disturbance.

In our experiment, we considered a simplified 3x3 unit cell, which has non-ideal truncation effects including asymmetry and reduced ground plane effects [fabricated samples for the case of the conventional standard dipole and the radio-transparent antenna are shown in Fig. 4(b)]. Additional details on each design and the testing methods are discussed in the Supplementary Materials. Figure 4 showcases the improvement in terms of far-field scattering offered by our radio-transparent antenna across the entire operating bandwidth (LB and HB). In particular, Fig. 4(c) compares the LB performance between the standard conductive dipole and the radio-transparent antenna, while Fig. 4(d) compares the HB performance in the three scenarios of Fig. 4(a), i.e., the conventional standard dipole, the radio-transparent antenna, and the isolated scenario (removing the LB element). The LB radiation performance supports very good matching between the standard dipole and the radio-transparent antenna patterns. For both considered frequencies, the beamwidth becomes slightly narrower in the radio-transparent antenna, due to its slightly thicker overall radius when including the dielectric. Yet, the comparison clearly shows good dipolar radiation from the radio-transparent antenna, essentially mimicking the performance of the conventional standard dipole.

Figs. 4(d) show the far-field performance at several frequencies across the HB. In all cases, the radiation from the radio-transparent antenna closely matches the one of the isolated panel, especially in terms of beam squint, gain and beamwidth. The standard LB dipole, on the contrary, produces a strong squint the HB beam, essentially re-directing the beam away from boresight. In fact, at boresight we see strong blockage by the standard dipole, reducing the HB gain by 4 dB across the band. The very low scattering of the radio-transparent antenna across the entire HB range is highly beneficial. We stress that our antennas are dual-polarized, and the far-field patterns in Fig. 4 demonstrate that the benefit of our radio-transparent antenna extends to both polarization planes.

The far-field patterns in Fig. 4(d) already demonstrate the improvements of the HB radiation patterns across a large bandwidth and angular spectra; however, in Fig. 5 we quantify the improvements offered by our approach using conventional performance metrics in commercial antenna systems. Beam squint and 3 dB beamwidth are the most challenging antenna metrics to restore when nearby obstacles introduce parasitic reflections. Up to the frequency of 1.91 GHz, both the standard dipole and the radio-transparent antenna have little impact on the pattern in terms of beam squint. However, above this frequency, the conventional standard dipole shows a strong deterioration on the HB radiation pattern. Yet, the radio-transparent antenna offers a remarkable field restoration across the whole band, with the squint becoming much lower and flatter. We emphasize that the standard dipole causes a re-direction of $20°-45°$ of the main beam between 2.310-2.710 GHz, while our cloaked antenna has a much lower range of re-directionality between $0°-5°$, except for a narrowband $10°$ squint at 2.610 GHz. Across the entire HB, the average beam squint caused by the standard dipole element was measured to be $14.1°\pm20.5°$, while the cloaked antenna average was $-0.45°\pm5.7°$. For comparison, the measured average beam squint of the isolated panel was $0.45°\pm6.8°$. It is also interesting to compare the measured beam squint to the total SCS calculated in Fig. 3, where we see the suppression bandwidth and the narrowband scattering peaks of the transparent dielectric core metasurface antenna. Below 2.0 GHz, the total SCS of the antenna increases, and a narrowband peak is noticed around 2.6 GHz. This peak is related to the angular stability of our design, which was minimized by using a simple inductive strip screen and is confirmed in our far-field measurements.

We have also studied the antenna gain and beamwidth, for which the standard dipole introduces strong beamwidth instability across the band with a reduced gain. The average gain in the presence of the LB standard dipole was measured to be $8.6\pm1.9$ dB. Meanwhile, the antenna cloak gain average was $11.9\pm2.5$ dB, and the average measure gain of the isolated case was $10.1\pm2.9$ dB. The average measured beamwidth for the standard dipole was $77.4°\pm14.0°$, radio-transparent dielectric antenna was $68.5°\pm7.2°$, and the isolated case was $72.2°\pm3.1°$. When considering the gain and beamwidth, it is important to consider that they are measured at the maximum beam angle. Therefore, we must consider that the radio-transparent antenna and isolated scenarios are measured nearly at boresight, while the standard dipole metrics are significantly skewed from by the beam squint they introduce. By comparing these performance metrics holistically, the field restoration of the radio-transparent dielectric antenna design is impressive.

In conclusion, we have introduced here a new concept to realize broadband and efficient radio-transparent antennas of critical importance for modern communication systems (3G, 4G, and 5G), as more and more antennas need to be integrated on the same platform with minimal mutual interference. In recently explored approaches to enable low-scattering antennas, a conductive cylindrical antenna forms the base of the radiating element, which is then covered with a metasurface to suppress its scattering over other frequency bands in which adjacent antennas operate. However, in these approaches the designed cloak typically works over a limited range of polarizations, incidence angles and frequencies. Here, we explored a conceptually new technique to enable broadband, dual-polarized radio-transparent antennas by exploiting the naturally low scattering of subwavelength dielectric objects. In this approach, the cloak not only suppresses undesired scattering from the dielectric core over a broad range of frequencies, but it also acts as an efficient radiator over the entire band of interest for transmit and receive operation. Such radio-transparent radiating elements can be of tremendous use in modern communication systems where more and more antennas need to be integrated into the same platforms. The proposed radio-transparent dielectric core metasurface antenna not only overcomes longstanding issues of conventional cloaking techniques in terms of narrow bandwidth, low angular stability, and sensitivity to the polarization of the incident wave, but it also shows that the cloak can act as an effective radiating element. Here, for proof of concept we applied this technique to antenna systems for 3G and 4G services; however, the same concept can be extended to 5G services that utilize several closely located frequency bands on the same platform, and to arbitrarily shaped antennas.

**References**

[1] C. A. Balanis, *Antenna Theory: Analysis and Design*. John Wiley & Sons, 2016.

[2] H.-H. Sun, B. Jones, Y. J. Guo, and Y. H. Lee, "Suppression of Cross-Band Scattering in Interleaved Dual-Band Cellular Base-Station Antenna Arrays," *IEEE Access*, vol. 8, pp. 222486-222495, 2020.

[3] N. Engheta, "Circuits with light at nanoscales: optical nanocircuits inspired by metamaterials," *Science*, vol. 317, no. 5845, pp. 1698-1702, 2007.

[4] N. Yu, P. Genevet, M. A. Kats, F. Aieta, J.-P. Tetienne, F. Capasso, and Z. Gaburro, "Light propagation with phase discontinuities: generalized laws of reflection and refraction," *Science*, vol. 334, no. 6054, pp. 333-337, 2011.

[5] N. Yu and F. Capasso, "Flat optics with designer metasurfaces," *Nature Materials*, vol. 13, no. 2, pp. 139-150, 2014.

[6] N. Meinzer, W. L. Barnes, and I. R. Hooper, "Plasmonic meta-atoms and metasurfaces," *Nature Photonics*, vol. 8, no. 12, p. 889, 2014.

[7] Y. Xu, Y. Fu, and H. Chen, "Planar gradient metamaterials," *Nature Reviews Materials*, vol. 1, no. 12, pp. 1-14, 2016.

[8] S. B. Glybovski, S. A. Tretyakov, P. A. Belov, Y. S. Kivshar, and C. R. Simovski, "Metasurfaces: From microwaves to visible," *Physics Reports*, vol. 634, pp. 1-72, 2016.

[9] Y. Zhao, A. N. Askarpour, L. Sun, J. Shi, X. Li, and A. Alù, "Chirality detection of enantiomers using twisted optical metamaterials," *Nature Communications*, vol. 8, no. 1, pp. 1-8, 2017.

[10] M. Chen, M. Kim, A. M. Wong, and G. V. Eleftheriades, "Huygens' metasurfaces from microwaves to optics: a review," *Nanophotonics*, vol. 7, no. 6, pp. 1207-1231, 2018.

[11] X. Tian, P. M. Lee, Y. J. Tan, T. L. Wu, H. Yao, M. Zhang, Z. Li, K. A. Ng, B. CK Tee, and J. S. Ho, "Wireless body sensor networks based on metamaterial textiles," *Nature Electronics*, vol. 2, no. 6, pp. 243-251, 2019.

[12] J. D. Binion, E. Lier, T. H. Hand, Z. Hao Jiang, and D. H. Werner, "A metamaterial-enabled design enhancing decades-old short backfire antenna technology for space applications," *Nature Communications*, vol. 10, no. 1, pp. 1-7, 2019.

[13] N. Mohammadi Estakhri, and A. Alù, "Wavefront Transformation with Gradient Metasurfaces," *Physical Review X*, Vol. 6, No. 4, 041008 (17 pages), October 14, 2016

[14] X. G. Zhang, W. X. Jiang, H. L. Jiang, Q. Wang, H.W. Tian, L. Bai, Z.J. Luo, S. Sun, Y. Luo, C. W. Qiu, and T. J. Cui, "An optically driven digital metasurface for programming electromagnetic functions," *Nature Electronics*, vol. 3, no. 3, pp.165-171, 2020.

[15] U. Leonhardt, "Optical conformal mapping," *Science*, vol. 312, no. 5781, pp. 1777-1780, 2006.

[16] J. B. Pendry, D. Schurig, and D. R. Smith, "Controlling electromagnetic fields," *Science*, vol. 312, no. 5781, pp. 1780-1782, 2006.

[17] A. Alù, and N. Engheta, "Achieving Transparency with Plasmonic and Metamaterial Coatings," *Physical Review E*, Vol. 72, No. 1, 016623 (9 pages), July 26, 2005

[18] D. Schurig, J. J. Mock, B. J. Justice, S. A. Cummer, J. B. Pendry, A. F. Starr, and D. R. Smith, "Metamaterial electromagnetic cloak at microwave frequencies," *Science*, vol. 314, no. 5801, pp. 977-980, 2006.

[19] W. Cai, U. K. Chettiar, A. V. Kildishev, and V. M. Shalaev, "Optical cloaking with metamaterials," *Nature Photonics*, vol. 1, no. 4, pp. 224-227, 2007.

[20] P. Alitalo and S. Tretyakov, "Electromagnetic cloaking with metamaterials," *Materials Today*, vol. 12, no. 3, pp. 22-29, 2009.

[21] R. Fleury, F. Monticone, and A. Alù, "Invisibility and cloaking: Origins, present, and future perspectives," *Physical Review Applied*, vol. 4, no. 3, p. 037001, 2015.

[22] H. Chu, Q. Li, B. Liu, J. Luo, S. Sun, Z. H. Hang, L. Zhou, and Y. Lai, "A hybrid invisibility cloak based on integration of transparent metasurfaces and zero-index materials," *Light: Science & Applications*, vol. 7, no. 1, pp. 1-8, 2018.

[23] P. S. Kildal, "Artificially soft and hard surfaces in electromagnetics," IEEE Trans. Antennas Propag., vol. 38, no. 10, pp. 1537-1544, 1990.

[24] M. Riel, Y- Bramd, Y. Demers, and P. de Maagt, "Performance improvements of center-fed reflector antennas using low scattering struts," *IEEE Trans. Antennas Propag.*, vol. 60, no. 3, pp. 1269-1280, March 2012.

[25] P. Alitalo, O. Luukkonen, L. Jylha, J. Venermo, and S. A. Tretyakov, "Transmission-line networks cloaking objects from electromagnetic fields," IEEE Transactions on Antennas and propagation, vol. 56, pp. 416–424, 2008.

[26] S. Tretyakov, P. Alitalo, O. Luukkonen, and C. Simovski, "Broadband electromagnetic cloaking of long cylindrical objects," Physical review letters, vol. 103, p. 103905, 2009.

[27] J. Vehmas, P. Alitalo, and S.A. Tretyakov, "Transmission-line cloak as an antenna," *IEEE Trans. Antenna and Wireless Prop. Lett.*, vol. 10, pp.1594–1597, 2011.

[28] A. Alù, "Mantle cloak: invisibility induced by a surface," *Phys. Rev. B*, vol. 80, 24115, 2009.

[29] P. Y. Chen and A. Alù, "Mantle cloaking using thin patterned metasurfaces," *Phys. Rev. B*, vol. 84, 205110, 2011.

[30] A. Monti, J. Soric, A. Alù, F. Bilotti, A. Toscano, and L. Vegni, "Overcoming mutual blockage between neighboring dipole antennas using a low-profile patterned metasurface," *Antennas and Prop. Lett.*, vol. 11, 1414, 2012.

[31] J. C. Soric, P.Y. Chen, A. Kerkhoff, D. Rainwater, K. Melin, and A. Alù, "Demonstration of an ultralow profile cloak for scattering suppression of a finite-length rod in free-space," *New J. Phys.*, vol. 15, pp. 033037, 2013.

[32] J. C. Soric, A. Monti, A. Toscano, F. Bilotti and A. Alù, "Multiband and wideband bilayer mantle cloaks," *IEEE Trans. Antennas Propag.*, vol. 63, no. 7, pp. 3232−3240, 2015.

[33] Z. H. Jiang, P.E. Sieber, L. Kang, and D.H. Werner, "Restoring intrinsic properties of electromagnetic radiators using ultralightweight integrated metasurface cloaks," *Adv. Funct. Mater.*, vol. 25, 4708−4716, 2015.

[34] H. M. Bernety, and A. B. Yakovlev, "Reduction of mutual coupling between neighboring strip dipole antennas using confocal elliptical metasurface cloaks," *IEEE Trans. Antennas Propag.*, vol. 63, pp. 1554-1563, 2015.

[35]     Y. R. Padooru, A. B. Yakovlev, P. Y. Chen, and A. Alù, "Line-Source Excitation of Realistic Conformal Metasurface Cloaks," *Journal of Applied Physics*, Vol. 112, No. 10, 104902 (11 pages), November 21, 2012.

[36]     A. Monti, J. Soric, A. Alù, A. Toscano, and F. Bilotti, "Design of Cloaked Yagi-Uda Antennas," *EPJ Applied Metamaterials*, Vol. 3, No. 10 (7 pages), November 23, 2016.

[37]     Y. R. Padooru, A. B. Yakovlev, P. Y. Chen, and A. Alù, "Analytical Modeling of Conformal Mantle Cloaks for Cylindrical Objects Using Sub-Wavelength Printed and Slotted Arrays," *Journal of Applied Physics*, Vol. 112, No. 3, 034907 (13 pages), August 14, 2012.

[38]     A. Forouzmand and A. B. Yakovlev, "Electromagnetic cloaking of a finite conducting wedge with a nanostructured graphene metasurface," *IEEE Trans. Antennas Propag.*, vol. 63, pp. 2191-2202, 2015.

[39]     H. M. Bernety, A. B. Yakovlev, H. G. Skinner, S. Y. Suh, and A. Alù, "Decoupling and Cloaking of Interleaved Phased Antenna Arrays Using Elliptical Metasurfaces," *IEEE Trans. Antennas Propag.*, Vol. 68, No. 6, pp. 4997-5002, June 2020.

[40]     G. Moreno, A. B. Yakovlev, H. M. Bernety, D. H. Werner, H. Xin, A. Monti, F. Bilotti, and A. Alù, "Wideband Elliptical Metasurface Cloaks in Printed Antenna Technology," *IEEE Trans. Antennas Propag.*, Vol. 66, No. 7, pp. 3512-3525, July 1, 2018.

[41]     T. V. Teperik and A. de Lustrac, "Electromagnetic cloak to restore the antenna radiation patterns affected by nearby scatter," *AIP Advances*, vol. 5, p. 127225, 2015.

[42]     S.A. Tretyakov, *Analytical Modeling in Applied Electromagnetics*. Norwood, MA: Artech House, 2003.

**Figures**

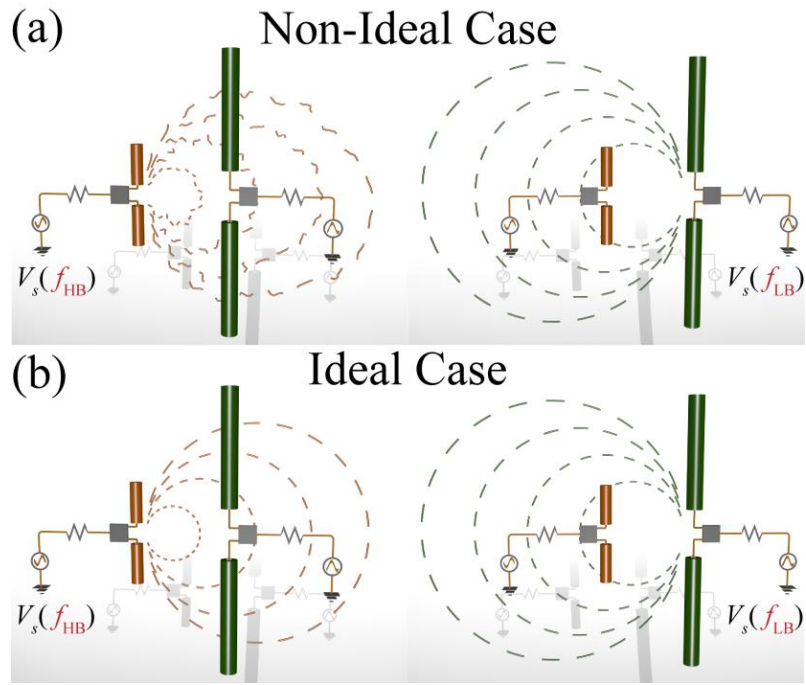

**Figure 1.** Schematic of closely located antennas: (a) Non-ideal scenario in which the low-frequency antenna drastically affects the radiation at high-frequencies; (b) Ideal scenario of a radio-transparent low-frequency antenna.

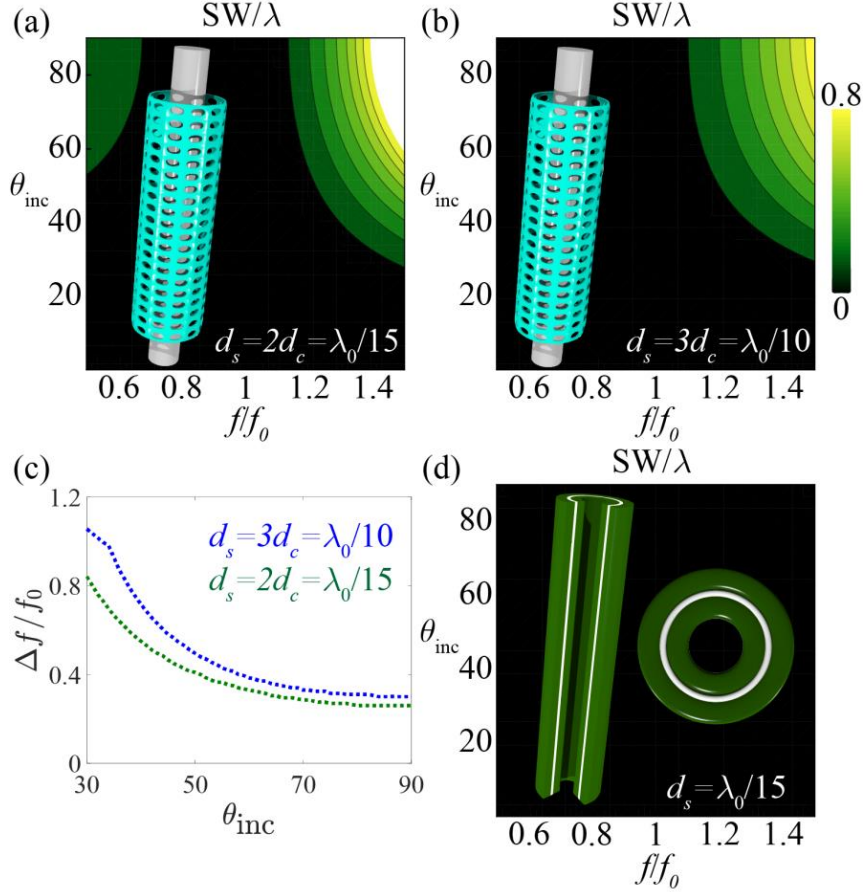

**Figure 2.** Comparison between scattering features of metasurface cloaks tailored to cloak conducting and dielectric cylinders. Panels (a) and (b) show the case of infinitely long conducting cylinders cloaked with an impedance metasurface. For (a), the radius of the impedance sheet is twice the one of the metallic cylinder $d_s = 2d_c = \lambda_0/15$ with air as the spacer, and the optimal sheet impedance of the cloak is $Z_s = -j0.28\eta_0$. For the design in panel (b) the radius of the impedance sheet is three times the one of the metallic cylinder $d_s = 3d_c = \lambda_0/10$ with air as the spacer, and the sheet impedance is $Z_s = -j0.54\eta_0$. Note that for the sake of this example, here we assume a ideal impedance sheets with no frequency dependence and the perforated screens shown in panels (a) and (b) are just illustrative schematics. Details of the calculation of the scattering width are given in the SoM. (c) Bandwidth comparison between the cases shown in (a) and (b). (d) Cloaking

of a hollow dielectric cylinder. For this design, the radius of the impedance sheet is $d_s = \lambda_0/15$, the thicknesses of super- and substrates are $\lambda_0/60$ and the sheet impedance is $Z_s = j1.39\eta_0$.

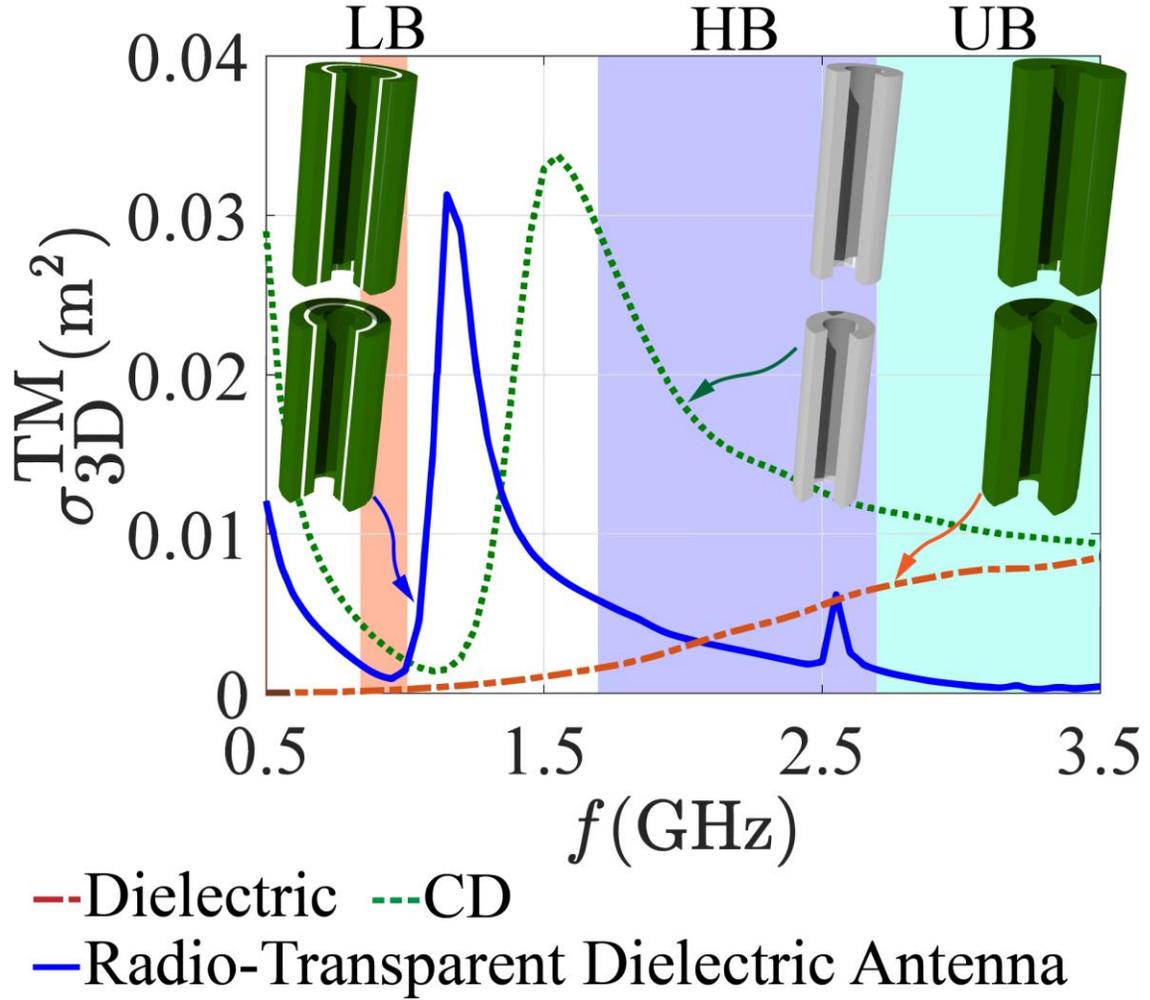

**Figure 3.** Total SCS for a normally incident plane wave across the bands of interest for a conventional conductive dipole (CD), a bare low-loss dielectric cylinder (without cloak) with $\varepsilon = \varepsilon_r(1 - j\tan\delta)$, where $\varepsilon_r = 4.4$ and $\tan\delta = 0.0005$, and for our radio-transparent antenna, all with same radii. The detailed dimensions and design details are given in Fig. S3 of the SoM.

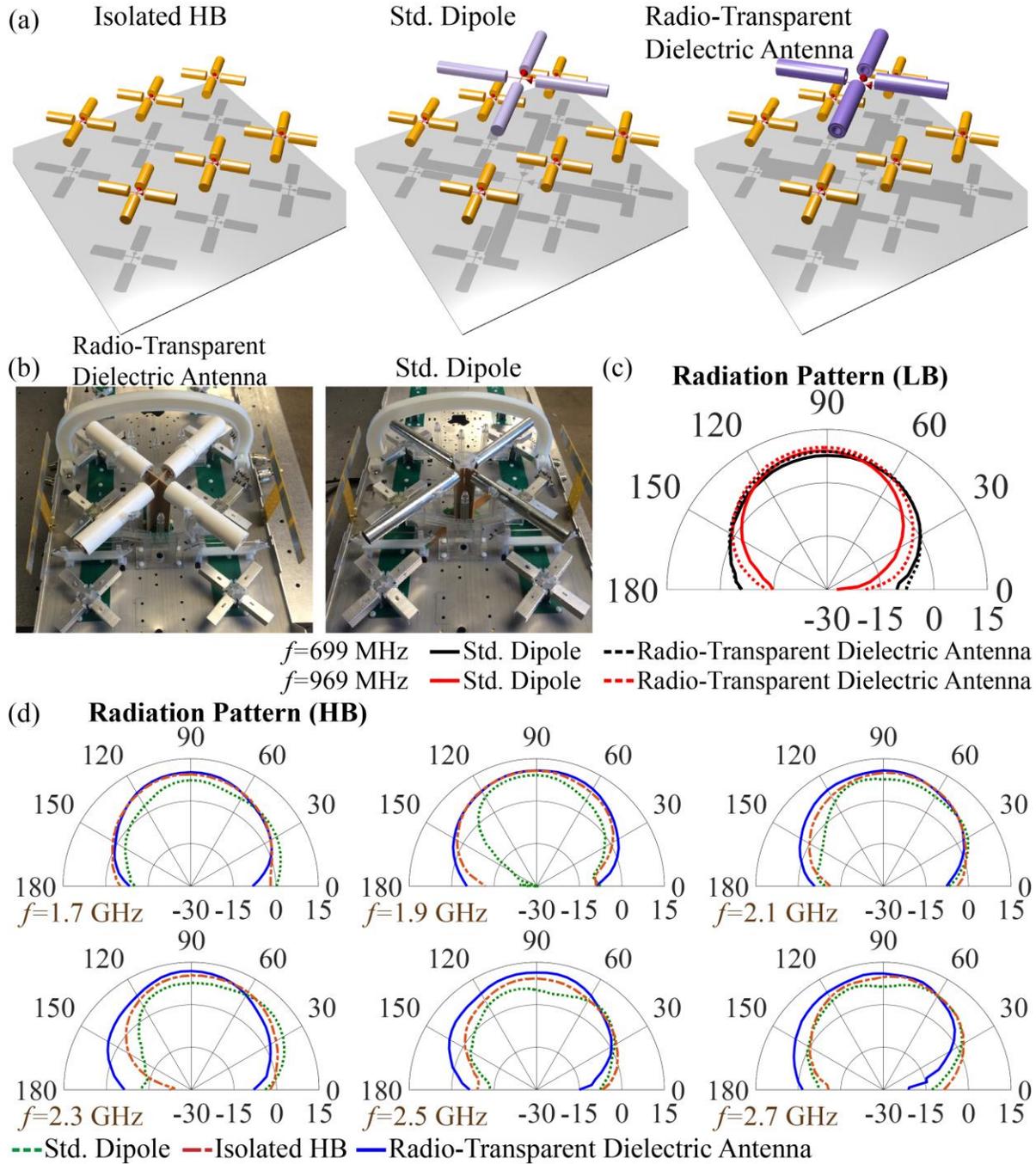

**Figure 4.** (a) Schematic of the isolated HB array, a typical communication system where a standard LB dipole is located over the HB array, and the proposed design where the radio-transparent antenna is located over the HB array. (b) Left panel: 3x3 panel unit cell of dual-polarized HB antenna elements with the proposed radio-transparent antenna above it. Right panel: Same unit

cell with a conventional CD LB element. Each HB panel is designed to operate between 1.710-2.710 GHz (45% bandwidth), while the LB elements operate from 698-968 MHz (32% bandwidth). Far-field comparison across the (c) LB and (d) HB (experimental results). The detailed dimensions of the designs presented in this figure are given in the SM.

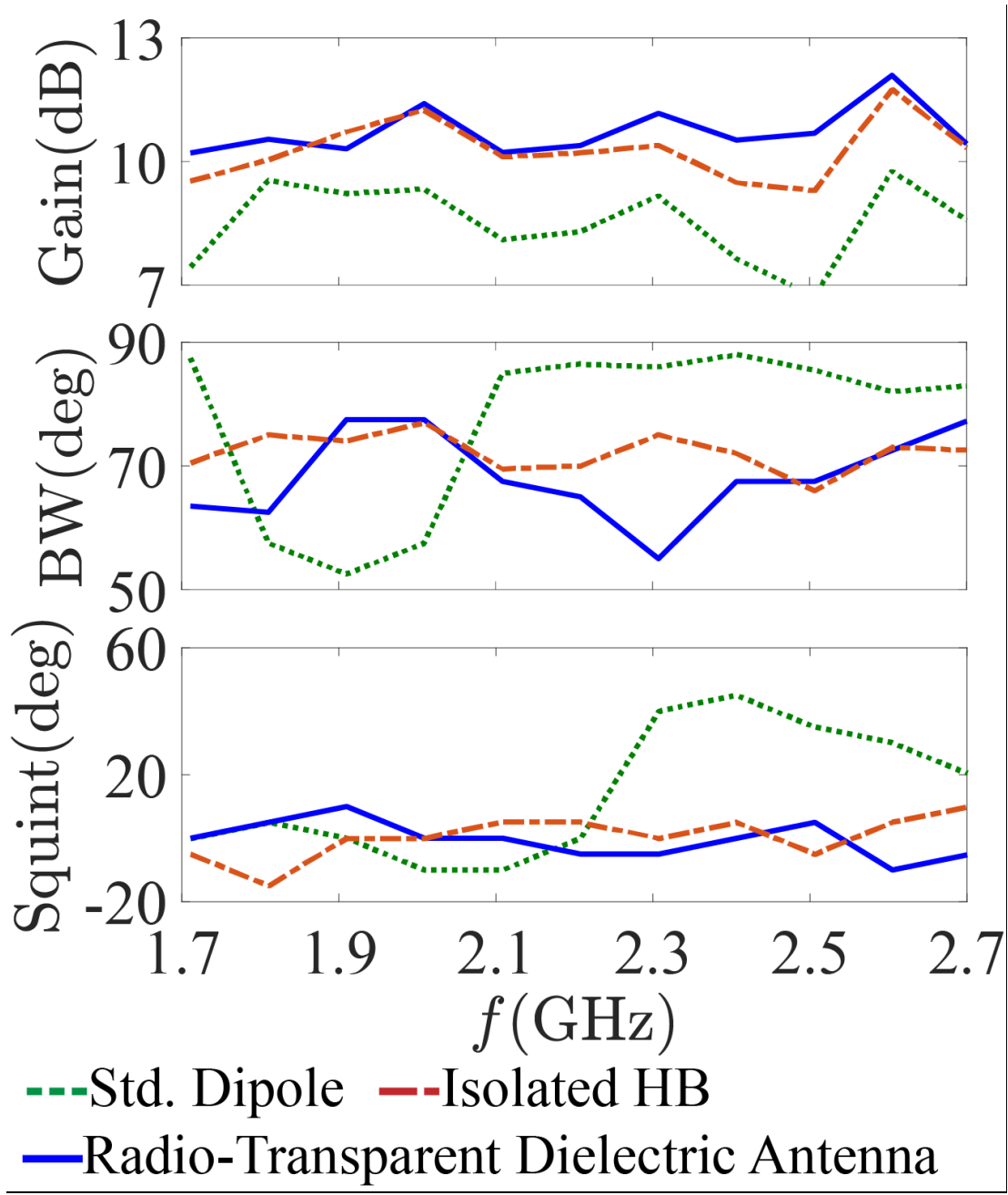

**Figure 5.** Radiation performance metrics for each testing case across the HB frequency range.

# Supplementary Material:

# Radio-Transparent Dipole Antenna Based on a Metasurface Cloak

Jason Soric[1], Younes Ra'di[2], Diego Farfan[1,2], and Andrea Alù[1,2,3,4]

[1]*Department of Electrical and Computer Engineering, The University of Texas at Austin, Austin, TX 78712, USA*
[2]*Photonics Initiative, Advanced Science Research Center, City University of New York, NY 10031, USA*
[3]*Physics Program, Graduate Center, City University of New York, NY 10016, USA*
[4]*Department of Electrical Engineering, City College of The City University of New York, NY 10031, USA*

*To whom correspondence should be addressed: aalu@gc.cuny.edu*

**Scattering from a metasurface integrated within a dielectric rod for improved angular stability and bandwidth suppression**

Eqs. (1)-(2) describe explicitly the fields in each region of the 2D cylinder under study, illustrated in the inset of Fig. S1 [1]:

$$E_z = E_{z0} e^{jk_{zi}z} \sum_n j^{-n} e^{jn\phi} \begin{cases} x_1 J_n(k_{\rho 1}\rho), & \rho < a_1 \\ x_2 J_n(k_{\rho 2}\rho) + x_3 Y_n(k_{\rho 2}\rho), & a_1 < \rho < a_2 \\ x_4 J_n(k_{\rho 3}\rho) + x_5 Y_n(k_{\rho 3}\rho), & a_2 < \rho < a_3 \\ J_n(k_{\rho 0}\rho) + c_n^{TM} H_n^{(2)}(k_{\rho 0}\rho), & \rho > a_3 \end{cases} \quad (1)$$

$$H_\phi = E_{z0} e^{jk_{zi}z} \sum_n j^{-(n+1)} e^{jn\phi} \begin{cases} \left(\dfrac{k_1}{k_{\rho 1}\eta_1}\right) x_1 J'_n(k_{\rho 1}\rho), & \rho < a_1 \\ \left(\dfrac{k_2}{k_{\rho 2}\eta_2}\right)\left[x_2 J'_n(k_{\rho 2}\rho) + x_3 Y'_n(k_{\rho 2}\rho)\right], & a_1 < \rho < a_2 \\ \left(\dfrac{k_3}{k_{\rho 3}\eta_3}\right)\left[x_4 J'_n(k_{\rho 3}\rho) + x_5 Y'_n(k_{\rho 3}\rho)\right], & a_2 < \rho < a_3 \\ \left(\dfrac{k_0}{k_{\rho 0}\eta_0}\right)\left[J'_n(k_{\rho 0}\rho) + c_n^{TM} H'^{(2)}_n(k_{\rho 0}\rho)\right], & \rho > a_3 \end{cases} \qquad (2)$$

Here $E_{z0} = E_0 \sin\theta_i$ is the incident electric field strength, $k_{zi} = k_0 \cos\theta_i$, $k_{\rho l} = \sqrt{k_l^2 - k_{zi}^2}$, and $\eta_l$ are the wavenumber components and wave impedance in each region $l$, respectively, where $l = 0$ corresponds to free space. In (1)-(2), $J_n(\xi)$ and $Y_n(\xi)$ are Bessel and Neumann functions of scattering order $n$, and the Hankel function is defined as $H_n^{(2)}(\xi) = J_n(\xi) - jY_n(\xi)$. In (2), $\psi'_n(\xi) = (d/d\xi)\psi_n(\xi)$. At each interface, we apply the continuity of the tangential fields and the impedance boundary condition, yielding a rank-6 determinant. While not considered in this work, the surface impedance is in general dyadic $\bar{\bar{Z}}(\phi) = Z_{zz}\hat{z}\hat{z} + Z_{z\phi}\hat{z}\hat{\phi} + Z_{\phi\phi}\hat{\phi}\hat{\phi} + Z_{\phi z}\hat{\phi}\hat{z}$, where $Z_{z\phi} = Z_{\phi z} = 0$ and $Z_{zz} = Z_{\phi\phi} \neq 0$ reduces to the isotropic case considered here, such that $E_z(a_2^+) = E(a_2^-) = Z_{zz}[H_\phi(a_2^+) - H_\phi(a_2^-)]$ [2].

Polarization coupling may be accounted for by considering the terms $\{Z_{z\phi}, Z_{\phi z}\}$, which may become detrimental or irrelevant depending on the cover topology. By design, simple scalar surface impedance covers show the lowest polarization coupling, which lead to well-behaved and large scattering suppression bandwidths. Recently, single and bi-layer surfaces have been employed to reduce total scattering to nearly zero, but at the cost of bandwidth reduction and potential polarization coupling between TM and TE modes [3]-[4]. Our goal here is to design the simplest cloaking structure that provides broadband and angle invariant scattering reduction.

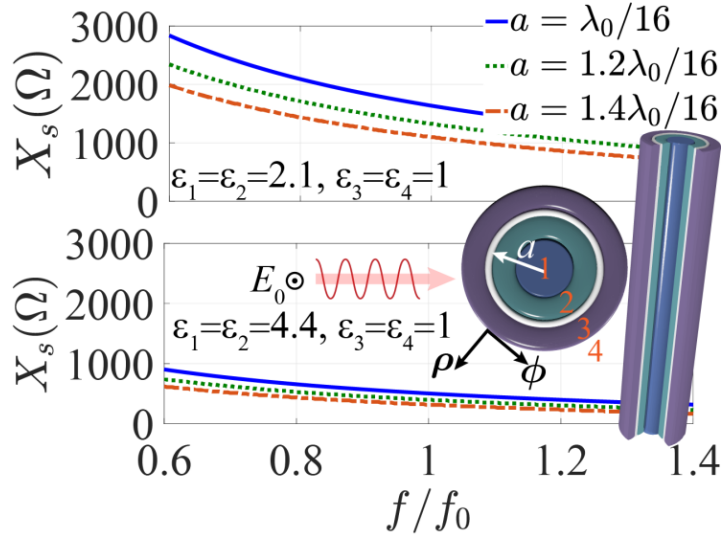

**Fig. S1.** Required surface impedance values from the minimization of scattering width (SW) at normal incidence for typical dielectrics with increasing cross-section dimensions.

The inset of Fig. S1 illustrates the cross-section of the cylindrical target under study. Here region 1 is the central region for $\rho < a_1$, region 2 is for $a_1 < \rho < a_2$, region 3 is for $a_2 < \rho < a_3$, and region 4 is free space for $\rho > a_3$. While the model is completely general, we only consider non-magnetic materials, such that, $\mu_l = \mu_0$, as they introduce unwanted passive intermodulation in high power sources such as base stations. It is interesting to first consider the required surface impedance across a wide frequency band for solid core dielectric rods of two readily available materials with a conformal cover at normal incidence to appreciate the required surface impedance with geometry and material composition. In Fig. S1, we increase the diameter of each rod to consider the required surface impedance to minimize the normalized total scattering width (SW), defined as

$$\sigma_{2D}/\lambda = \frac{2}{\pi \sin\theta_i} \sum_{n=0}^{N_{max}} (2-\delta_{0n})|c_n^{TM}|^2$$ where $\delta_{0n}$ is the Kronecker discrete delta function, $\lambda$ is the free-space wavelength, and $N_{max}$ is the maximum relevant order. We safely consider the ultrathin conductive impedance layer as being lossless and isotropic, such that, $Z_{zz} = jX_s$. Figure S1 illustrates two important features of the choice of realizable scalar impedance surfaces. Employing the scattering cancellation technique practically requires moderate scattering; therefore, electrically small targets of low dielectric composition have strong frequency dispersion across the band and also require large effective surface impedances. Therefore, in a practical design for large bandwidth performance, dielectric targets of moderate electrical cross-section should be considered. Care must also be taken to not introduce TE-polarized scattering by choosing too large the electrical cross-section, which in most designs can be 20 dB lower than that of the dominant TM-polarized wavefronts for the frequencies considered here. Said explicitly, we do not aim to reduce both polarizations, but significantly reduce the dominant mode, while not enhancing the already much lower higher-order modes.

Our approach is based on the reduction of scattering from moderate-valued dielectric rods of low electrical cross-section [5]. By choosing this route, we leverage several features previously unexplored. The scattering of dielectric rods is already much less than comparable conductive targets at all angles and polarizations; therefore, simple scalar covers targeting the dominant TM-pol are appropriate. Additionally, conformal covers optimize the bandwidth for dielectric targets, which is diametric to the bulky covers targeting conductive targets [6]-[7], or with high dielectric substrates [8]. The simplest inductive surface for dominant TM-polarization is that of traces aligned with the incident wavefront [9]

$$Z_{strips}^{TM} = j\omega \frac{\mu_0 D}{2\pi} \ln\left[\csc\left(\frac{\pi w}{2D}\right)\right]\left(1 - \frac{\cos^2 \theta_i}{2\varepsilon_{eff}}\right). \tag{3}$$

This simple but accurate formula calculates the effective shunt inductance for a given infinite planar interface, with trace width $w$, period $D$, and $\varepsilon_{eff} = (\varepsilon_l + \varepsilon_{l-1})/2\varepsilon_0$. It is well-known that thin angular stable surfaces for various applications are difficult to design while maintaining their functionality [9]-[11], and this intrinsic angular sensitivity is clearly seen in (3) as well. Different from cloaking applications in the strict sense [5], here we immerse the inductive screen inside the target rather than covering it to improve the angular stability and suppression bandwidth. In Fig. S2, we demonstrate the improvements in angular stability and bandwidth across a large frequency band. The analytical SW in Fig. S2 includes the elevation angular dependence for realizable simple covers as well as their frequency dispersion using (3). We emphasize two important considerations here. First, this model does not include polarization coupling effects, due to the anisotropy of realistic covers, which may affect the angular performance. Second, these analytical results are for a 2D rod, which does not include longitudinal resonances, end effects, and relevant to this work, the input impedance at the feedgap when formed into an antenna. Still, these analytical results give us great insight into the effect of surface immersion suggested in this work.

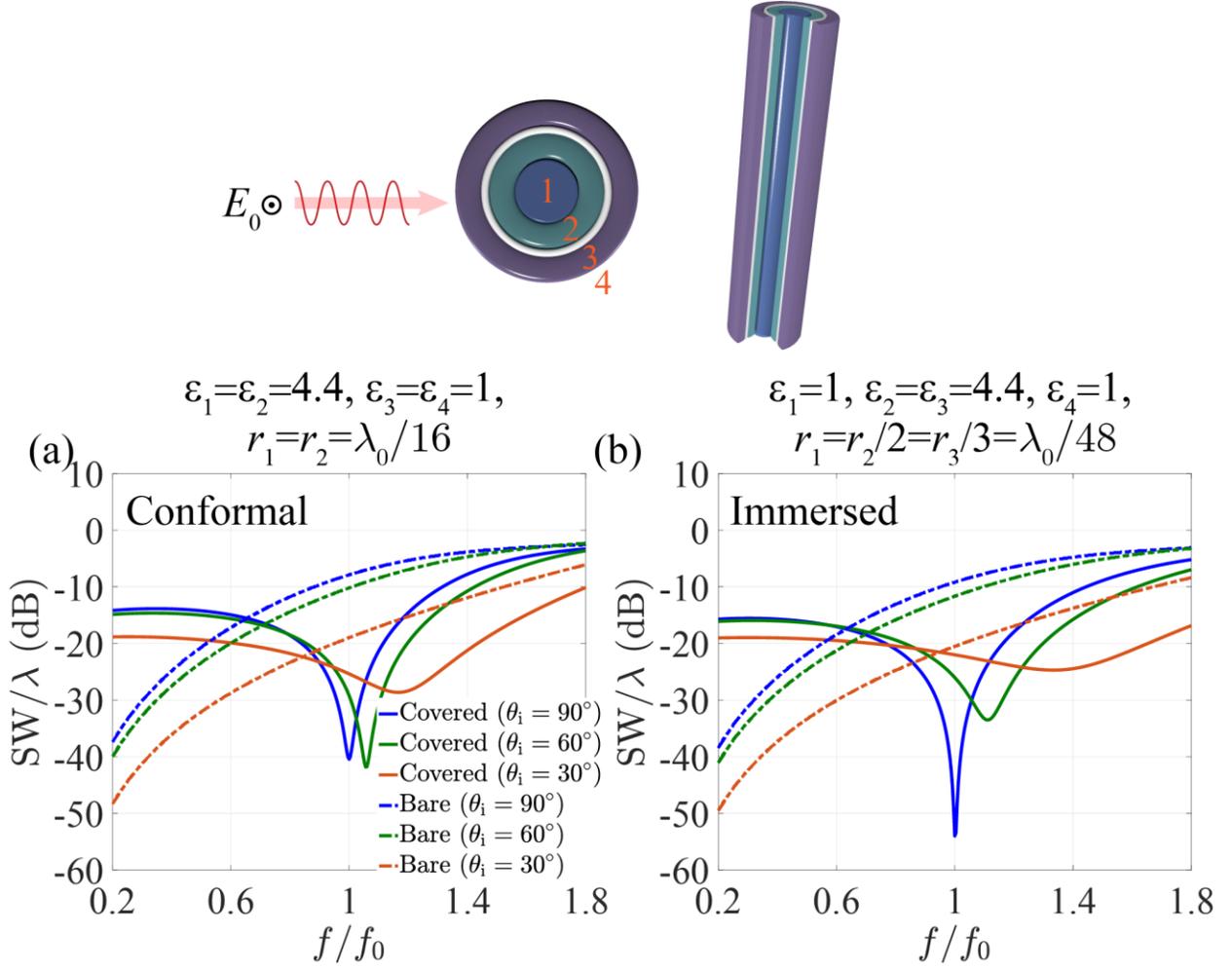

**Fig. S2.** SW with incident angle for conformal and immersed covers.

Figure S2 considers the angular SW response across the band of interest for different incident angles. Here, we choose the same surface impedance calculated by (1)-(2) for a given frequency at normal incidence and then use (3) to take into account frequency dispersion and angular dependence of a practical impedance surface. For the conformal design, the dielectric rod is solid with $\varepsilon_l = \varepsilon_c = 4.4$ and $d = 21.9 mm \approx 1.4 d_0$. A surface impedance of $j1.34\eta_0$ (where $\eta_0$ is freespace wave impedance) is applied to the surface, as calculated for normal incidence for $f/f_0 = 1$. As it can be seen from the figure, there is a blue shift in the minimum scattering dip as the incident angle increases which is fully consistent with the experimental work in Ref. [12].

Next, we consider an immersed mantle cloak design. The total width of both designs is the same. For the immersed case, the dielectric is inhomogeneous, where the center has been hollowed out for antenna mounting and weight reduction, as well as capacitive feeding for passive intermodulation reduction. The improvement in angular stability and suppression bandwidth is significant in the immersed case. As well, the overall visibility of the sample is now much lower across the entire band, consistently being better than that of the conformal design. Now the required surface impedance is $j1.1\eta_0$ at $f/f_0 = 1$ for normal incidence.

**Dual-polarized spectral and angular response for the designed TDCMA**

To verify the angular response across a large bandwidth, we excite the TDCMA designed for Fig. 3 of the main text with dual-polarized plane waves along the elevation plane. The optimized antenna has a trace width of $w$ with a period $D = a_2/\pi$. In Fig. S3, we still demonstrate that this design gives very good bistatic scattering suppression, and more impressively, with an angular stable response over a very broad bandwidth. One obvious discrepancy, is now the TE-polarized wavefronts introduce scattering beyond that of the conductive dipole (CD) for oblique incidence. While this is important for dual-polarized antennas, there are two main consequences of this. First, this TE scattering enhancement simply limits the bandwidth for dual-polarized sources; however, the bandwidth is still remarkable and well-above most other cloaking strategies. Second, this simply implies this particular design is best suited for moderate to high gain systems. For the system considered in this work, the gain of the reflector-backed array is high enough to not be drastically affected by large off-angle radiation. We note that the elevation angle in this study has been redefined to consider the beamwidth of incident radiation [see Fig. S3(a)]. Considering the specification of both the LB and HB BW to be < 70 deg., we expect significant radiation for $\theta_i < 30°$.

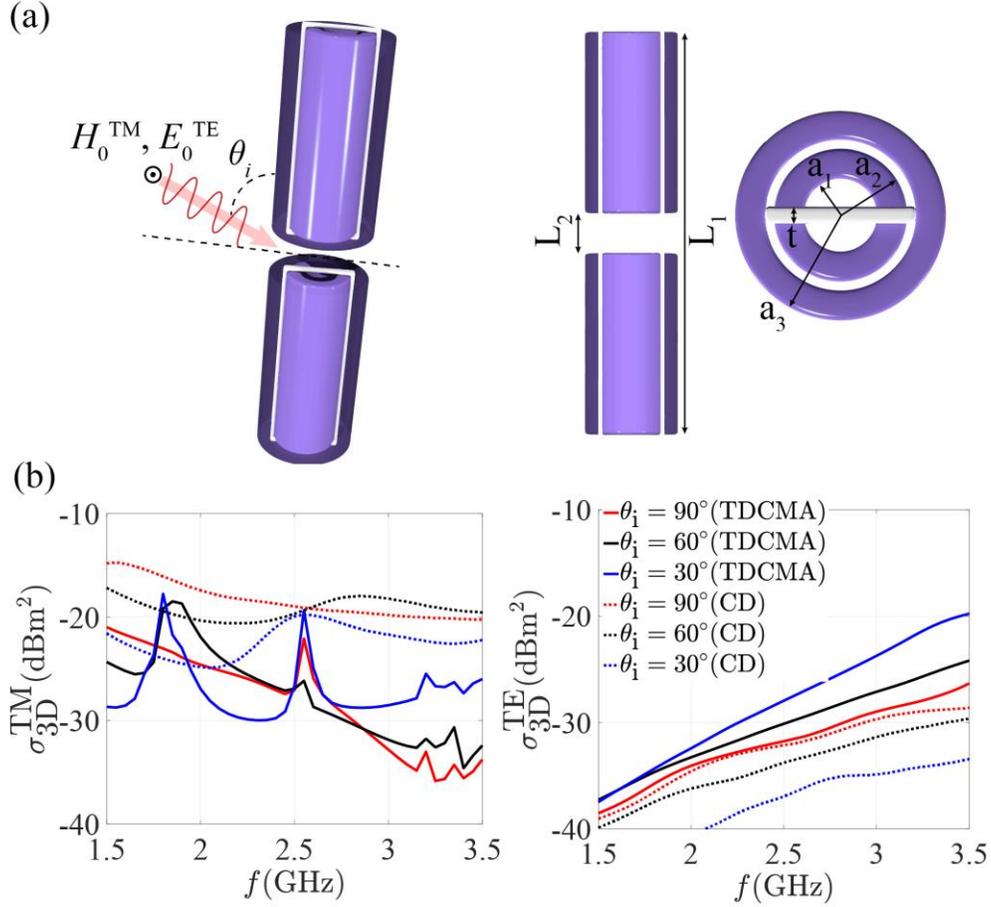

**Fig. S3.** Dual-polarized spectral and angular response for the designed TDCMA. Design parameters: $L_1 = 231\text{mm}$, $L_2 = 36.4\text{mm}$, $a_1 = 4.77\text{mm}$, $a_2 = 7.77\text{mm}$, $a_3 = 10.77\text{mm}$, and $t = 1.5\text{mm}$. These dimensions are the same for the design in Fig. 3 of the main text. Simulations have been performed using CST STUDIO SUITE [13].

We note several narrowband scattering peaks across the band at 1.74, 2.5, and 3.25 GHz. These effects are caused by the realistic anisotropy of the cover and longitudinal resonances of the finite length antenna. Again, this points to the necessity of using a simplified surface for broadband applications with angular rich near-field sources.

**Design of a simple cloak for practical antenna applications**

As discussed previously, large bistatic suppression bandwidths and angular spectra are already difficult design challenges in their own right. An equally important consideration is that the low-observable dipole arms, when formed into an antenna, radiate with industry-level standards. These lowband (LB) specifications require that the antenna have a return loss better than 10 dB when matched to a standard 50 $\Omega$ system; the 3 dB beamwidth $(BW) < 70°$, and beam squint $< \pm 7°$, thus, directing basestation links efficiently into predefined sectors.

To demonstrate the improvements offered by our design, we test this proposed cloaking method with a unit cell of a real basestation antenna panel. In Fig. S4, we show the geometry of this single unit cell. Placed about a quarter of the central wavelength (at 2.2 GHz) above the panel backplane are six dual-polarized highband (HB) elements. A large LB cross-dipole is then placed about a quarter of its central wavelength (at 833 MHz) above the same backplane. Due to the close proximity of the HB and LB elements, a very low-profile and conformal design is needed, which is difficult to obtain with bulky covers designed to reduce the scattering of conductive targets over a large bandwidth in conventional approaches, in this case, a conventional cylindrical dipole [6]-[7]. Detailed design dimensions are given in Figs. S4(b) and S4(c). The fabricated panel is shown in Fig. S4(d).

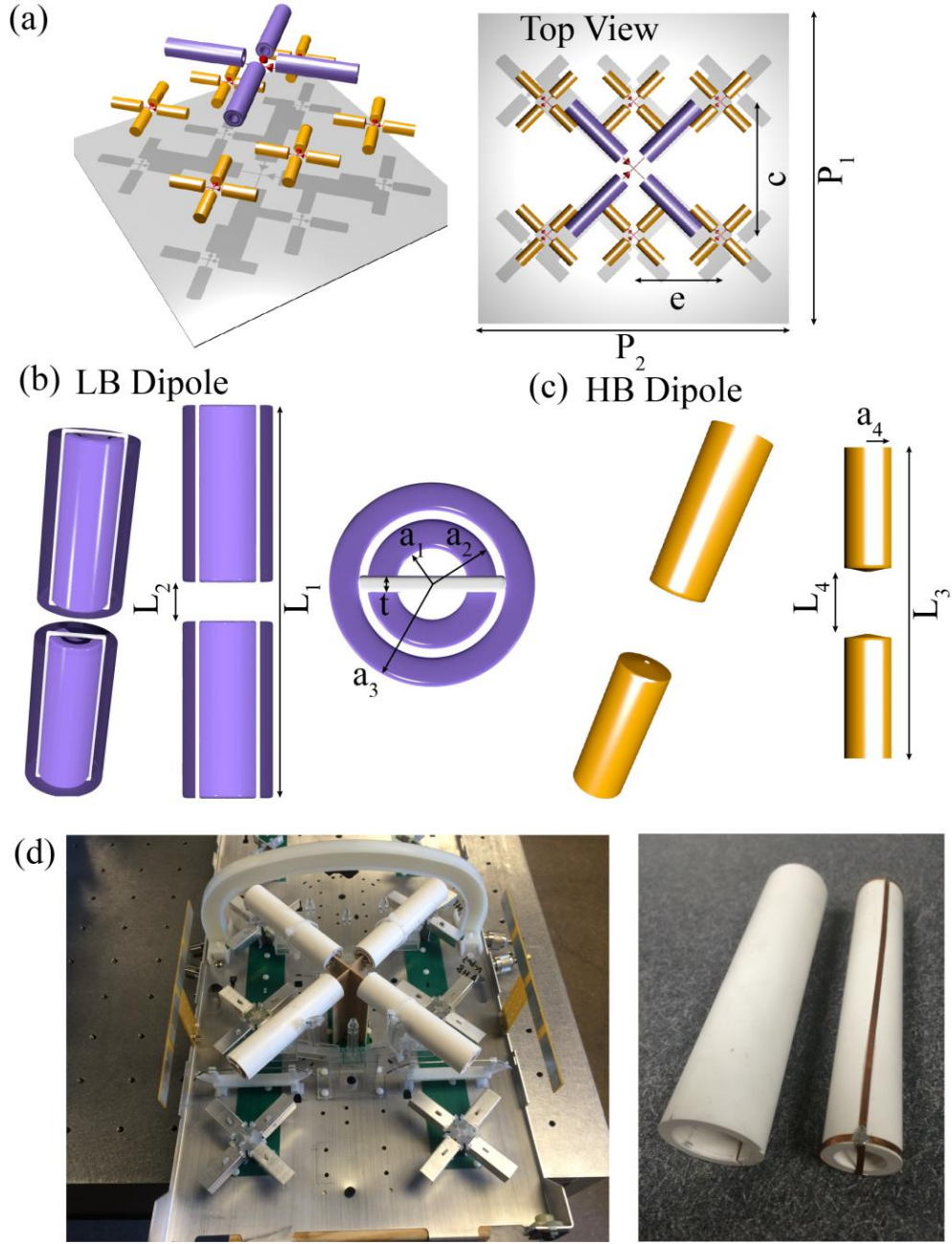

**Fig. S4.** (a) Geometry of the single panel. (b) A LB TDCMA arm. (c) A HB TDCMA arm. (d) Fabricated panel. Design parameters: $P_1 = P_2 = 352$mm, $e = 112$mm, $c = 172$mm, $L_1 = 231$mm, $L_2 = 36.4$mm, $L_3 = 100$mm, $L_4 = 18$mm, $a_1 = 4.77$mm, $a_2 = 7.79$mm, $a_3 = 11.44$mm, $a_4 = 7.54$mm, and $t = 1.5$mm. The distances between the ground plan and center of the HB array TDCMA antenna are 38mm and 83.3mm, respectively.

**Lowband performance: Cloaking surface radiator**

To use the cloaking surface itself as a good radiator, it must support pure dipolar radiation without beam-squinting. Twisted double helix designs were first suggested in the conference paper [14] to meet the large surface impedance needed; however, these covers suffered from narrow LB matching performance, beam squint, and cross-polarization. The simple design in this work has much better polarization purity giving squint-less dipolar radiation with excellent beamwidth. The main shortcoming of the design in Ref. [14] was its similarity to a normal mode helix, which are inherently narrowband antennas [15]. The design presented here is much easier to match across a large bandwidth, using a planar stub and a $\lambda/4$ transformer, implemented on a 3-layer board as is typically done in such antenna systems. Fine-tuning of the matching and beamwidth was also made by adjusting the feedgap and distance to the reflector.

In Fig. 4(c) of the main text, we have shown that the radiation patterns at the LB edges (699 and 969 MHz) match the conventional CD very well in terms of antenna gain, squint and BW, and we will show even more comparisons later in this Supplementary Material. In Fig. S5 we compare the matching performance measured from the two cross-dipoles in Fig. S4(d). The specified return loss of 10 dB is matched from 659-997 MHz, which exceeds the LB specification. The matching circuit was made for a single dipole, but when constructed into a cross-dipole, a narrowband spike at 914 MHz was observed with level -6.7 dB. While this spike is very narrow, it most likely due to near-field coupling between the two polarizations. Even still, the 10 dB LB bandwidth is 36%, which is quite good especially considering the low complexity integrated matching circuit and the HB transparency of the cross-dipole.

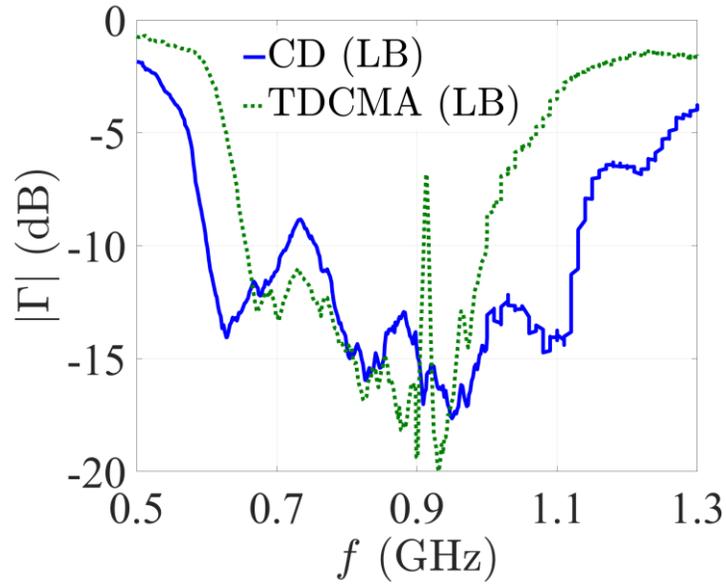

**Fig. S5.** Matching performance of the proposed and conventional LB antennas.

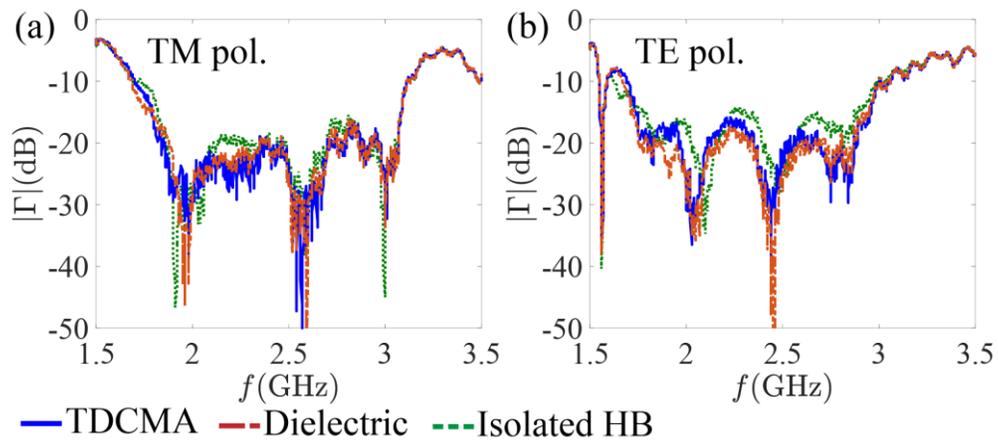

**Fig. S6.** HB matching comparison.

**High-band matching performance**

Figure S6 compares the matching of the highband elements without any LB element (Isolated HB), the TDCMA, and the CD mounted in front of them. Here, we see both blockers have only a marginal effect on the matching of the highband elements. However, for the case of the TE-pol, some noticeable improvement is shown by the TDCMA replacing the CD. We do note that for both polarizations, the TDCMA does match the Isolated HB cases better across the HB.

**Near-field verification for dominant TM-pol**

While not shown in the main paper, it is interesting to consider the near-field effects of each of the LB antenna elements. Here we consider only the TM-pol because of its dominant scattering characteristics. The details of our in-house near-field scanner can be found in Ref. [12]. For the following, the measurement raster scan area is taken over an area of 396×396 mm at a height of $z = 173$mm (3/4 the CD dipole length). The exciter in this case is a 5 dBi log-periodic antenna, which is the same antenna used as the receiver in the far-field measurements. Below in Fig. S7, we plot the near-field scattering gain (SG) with background subtraction. This near-field figure of merit shows the difference in scattering between the TDCMA and the CD. Remarkably, we see in-plane bistatic scattering suppression for over 3 GHz, most of which being below 6 dB. Compared to the cloak in Ref. [7], the increase in suppression bandwidth here is dramatic. We highlight five notable frequencies to compare the near-field snapshots-in-time in Fig. S8 for impinging wavefronts polarized along **z** and propagating along **y**.

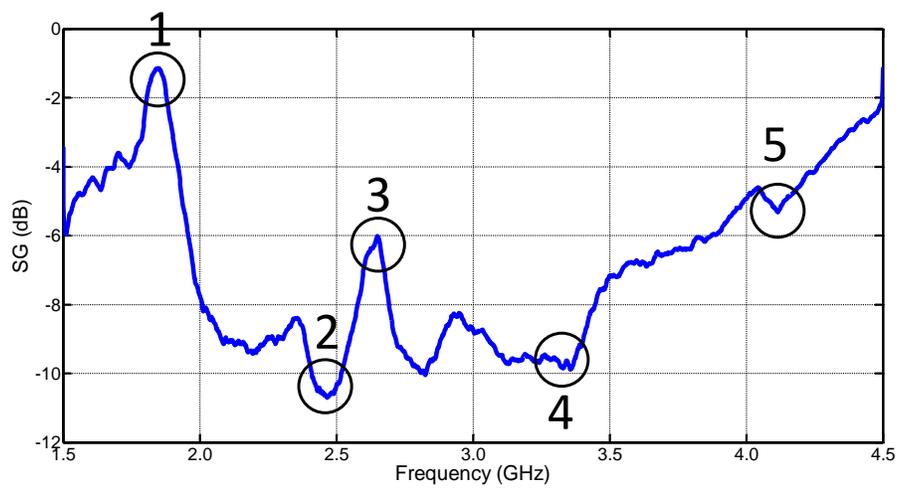

**Fig. S7.** Near-field scattering gain.

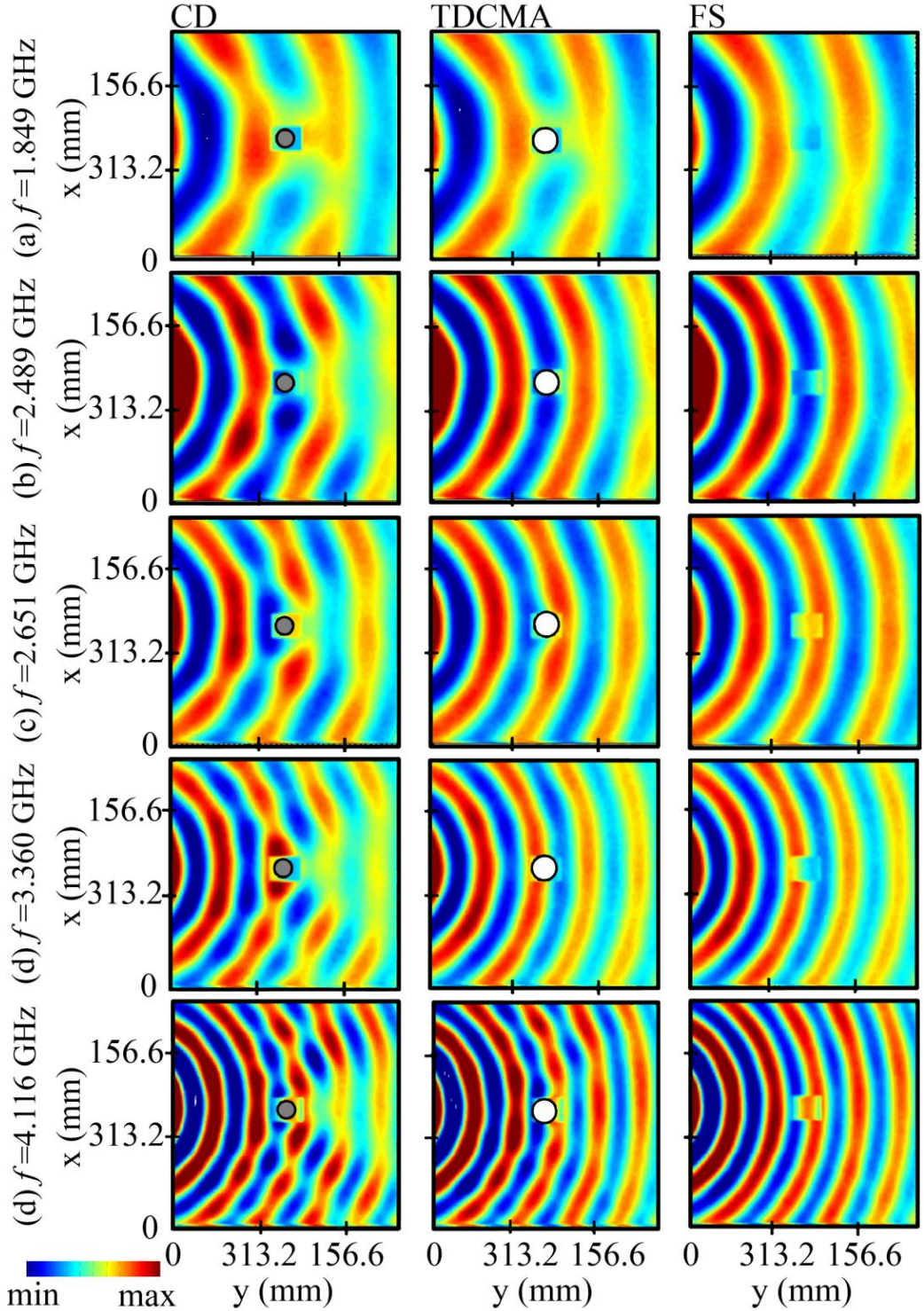

**Fig. S8.** Snapshots in time of the total electric field at different frequency points for conductive dipole, TDCMA, and free space cases (conductive dipole and TDCMA cross-section to relative scale). All units are in mm.

**Far-field measurement details**

Figure S9 illustrates the setup of the far-field measurements of the main text. Here the unit cell is a 3×2 array of HB cross-dipoles, and the far-field measurements are compared by removing all LB elements (Isolated HB) and then mounting either the CD or TDCMA, and taking measurements for all three cases. The receiver (Rx) in each of these measurements is a broadband log-periodic antenna with approximately 5 dBi gain across the bands of interest, and the unit cell panel is the transmitter (Tx) in each case. The distance between the Rx and Tx is $R = 1.25$m to be sufficiently in the far-field across the bands, according to $2l^2/\lambda$ [15], where $l$ is the longest dipole length and $\lambda$ is the wavelength at the highest frequency.

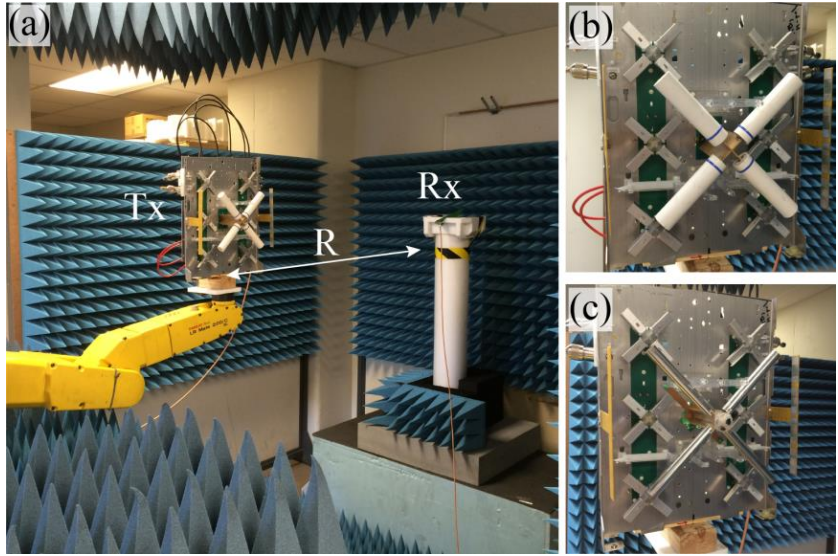

**Fig. S9.** (a) Antenna gain measurements for the low and high-bands. (b) TDCMA unit cell. (c) CD unit cell.

To measure the gain in each scenario, we rotate the antenna panel using our robot, as seen in Fig. S9(a), and calculate $G_{Tx} = |S_{21}|^2 / \{G_{Rx}[\lambda/(4\pi R)]^2$, where $S_{21}$ is the transmission measured from a vector network analyzer, and $G_{Rx}$ is gain of the receiver [15]. The gain of the receiver was first measured by using a pair of log-periodic antennas prior to the test shown in Fig. S9(a). In Fig. S10, we show frequency samples not shown in the main paper to fully appreciate the HB antenna restoration by the TDCMA approach.

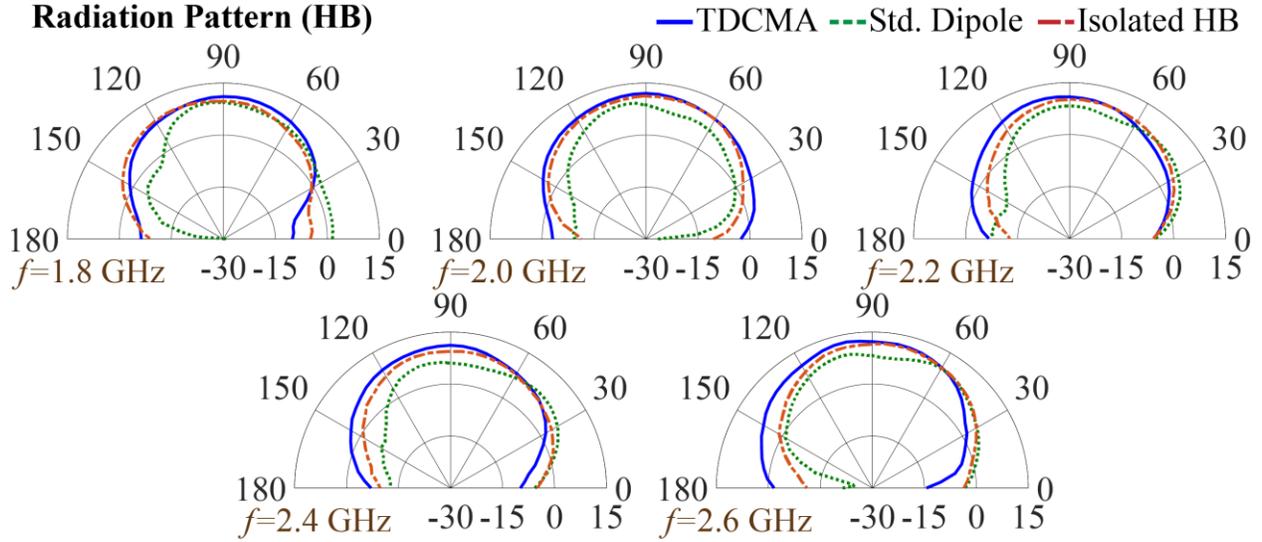

**Fig. S10.** Measured far-field results across HB with each blocker, or of the exciter panel only.

**References**

[1] C.A. Balanis, *Adv. Engineering Electromagnetics 3rd*, John Wiley & Sons, Hoboken, NJ, 1989.
[2] T.B.A. Senior and J.L. Volakis, *Approximate Boundary Conditions in Electromagnetics*, IEEE, London, 1995.
[3] Z.H. Jiang and D.H. Werner, "Exploiting metasurface anisotropy for achieving near-perfect low-profile cloaks beyond the quasi-static limit", *J. Phys. D: Appl. Phys.*, vol. 46, 505306, Nov. 2013.
[4] J.C. Soric, A. Monti, A. Toscano, F. Bilotti, and A. Alù, "Multiband and wideband bilayer mantle cloaks", *IEEE Trans. Antennas and Propaga.*, vol. 63, no. 7, 3235−3240, July 2015.
[5] A. Alù, "Mantle cloak: invisibility induced by a surface," *Phys. Rev. B*, vol. 80, 24115, 2009.

[6] Z.H. Jiang, P.E. Sieber, L. Kang and D.H. Werner, "Restoring intrinsic properties of electromagnetic radiators using ultralightweight integrated metasurface cloaks", *Adv. Funct. Mater.*, vol. 25, no. 29, 4708−4716, June 2015.

[7] J.C. Soric, A. Monti, A. Toscano, F. Bilotti, and A. Alù, "Dual-polarized reduction of dipole antenna blockage using mantle cloaks", *IEEE Trans. Antennas and Propagat.*, vol. 63, no. 11, 4827−4834, Sept. 2015.

[8] A. Monti, J.C. Soric, A. Alù, F. Bilotti, A. Toscano, and L. Vegni, "Overcoming mutual blockage between neighboring dipole antennas using a low-profile patterned metasurface", *IEEE Antennas and Propaga. Lett.*, vol. 11, 1414−1417, Nov. 2012.

[9] O. Luukkonen, C. Simovski, G. Granet, G. Goussetis, D. Lioubtchenko, A. V. Raisanen, and S. A. Tretyakov, "Simple and accurate analytical model of planar grids and high-impedance surfaces comprising metal strips or patches," *IEEE Trans. Antennas Propagat.*, vol. 56, pp. 1624−1632, 2008.

[10] E.F. Knott, J.F. Shaeffer, and M.L. Tuley, *Radar Cross Section: its prediction, measurement, and reduction*, Artech House, MA, 1985.

[11] B.A. Munk, *Frequency Selective Surfaces: Theory and Design*, Wiley, New York, 2000.

[12] J.C. Soric, P.Y. Chen, A. Kerkhoff, D. Rainwater, K. Melin, and A. Alù, "Demonstration of an ultralow profile cloak for scattering suppression of a finite-length rod in free-space," *New J. Phys.*, vol. 15, 033037, March 2013.

[13] CST STUDIO SUITE 2018, https://www.3ds.com/products-services/simulia/products/cst-studio-suite/.

[14] J.C. Soric and A. Alù, "Radio-frequency transparent dipole antennas," *IEEE AP-S/URSI International Symposium*, 53, July 2015, Vancouver, BC.

[15] C.A. Balanis, Antenna Theory: Analysis and Design 3rd Ed. Hoboken, NJ: Wiley, 2005.